\documentclass{aastex631}
\usepackage{aas_macros}
\usepackage{amsmath}
\usepackage{physics}
\usepackage{graphicx}
\usepackage{xcolor}
\usepackage{float}
\usepackage{algorithm,algpseudocode}
\usepackage{booktabs}

\begin{document}

\author[0009-0005-5941-1845]{Johannes Martin}
\affiliation{Eindhoven University of Technology, The Netherlands}
\author[0000-0001-6338-9728]{Jeremiah Lübke}
\affiliation{Institut für Theoretsiche Physik I, Ruhr-Universität Bochum, Germany}
\author[0000-0002-1057-8565]{Tianyi Li}
\affiliation{Dept. of Physics and INFN, University of Rome Tor Vergata, Rome, Italy}
\author[0000-0002-7162-5038]{Michele Buzzicotti}
\affiliation{Dept. of Physics and INFN, University of Rome Tor Vergata, Rome, Italy}
\author[0000-0003-0622-071X]{Rainer Grauer}
\affiliation{Institut für Theoretsiche Physik I, Ruhr-Universität Bochum, Germany}
\author[0000-0001-8767-9092]{Luca Biferale}
\affiliation{Dept. of Physics and INFN, University of Rome Tor Vergata, Rome, Italy}

\date{\today}

\title{Generation of cosmic ray trajectories by a Diffusion Model trained on test particles in 3D magnetohydrodynamic turbulence}

\begin{abstract}
Models for the transport of high energy charged particles through strong magnetic turbulence play a key role in space and astrophysical studies, such as describing the propagation of solar energetic particles and high energy cosmic rays.
Inspired by the recent advances in high-performance machine learning techniques, we investigate the application of generative diffusion models to synthesizing test particle trajectories obtained from a turbulent magnetohydrodynamics simulation.
We consider velocity increment, spatial transport and curvature statistics, and find excellent agreement with the baseline trajectories for fixed particle energies.
Additionally, we consider two synthetic turbulence models for comparison.
Finally, challenges towards an application-ready transport model based on our approach are discussed.
\end{abstract}


\section{Introduction}

The study of fast particle propagation in turbulent magnetic fields is a fascinating area of research with many open questions. An accurate description of cosmic ray transport is crucial for understanding the signatures of cosmic ray electrons and hadrons, as well as gamma rays and neutrinos. Turbulence plays a key role in governing particle transport in galactic outflows and the interstellar medium, which is typically modelled using diffusion tensors in transport equations. In recent years, the significance of intermittency, coherent structures, and local mirror configurations within underlying magnetohydrodynamic turbulence has become evident for understanding the complex diffusion process (see e.g.~\citet{sampson2022streamingcrdiffusion,reichherzer2023efficientmicromirrorconfinementsubtev,ewart2024radiobubbleconfinement,butsky2024intermittent,barretomota2024cosmicraydiffusionturbulent,Youssef2024cosmicraydiffusionintwolocalfilamentaryclouds,ntormousi2024strongturbulencemagneticcoherent}).
These insights highlight the need for new approaches to address this topic effectively.

A natural approach to the study of particle transport in turbulence consists in a numerical solution of the magnetohydrodynamic (MHD) equations and the propagation of test particles in turbulent MHD fields. This approach has been applied to simulate small regions of interest to extract processes and effective diffusion coefficients as input for coarse-grained transport descriptions. The dream to describe significantly larger regions of interest (e.g., the heliosphere, see below, the interstellar medium, or galactic outflows) in this way will, however, remain one for the foreseeable future: for simulations covering, e.g., the whole heliosphere one has to resolve small scales at the ion-gyro radius $(\sim 10^{5} \mathrm{~m})$, where damping processes operate, up to about $100 \mathrm{AU}$, which is the distance to
the solar wind termination shock. This would result in a mesh of about $1\,500\,000^{3}$ points,
a size which is not possible to handle in the near future. 

To circumvent this problem, different strategies involving the generation of synthetic turbulence fields have been designed, as described below. However, all the methods presented so far have shortcomings that may affect simulation results to different degrees. For example, most methods only require as input the shape of the energy spectrum of the synthetic turbulence (e.g. that of a Kolmogorov spectrum \citep{kolmogorov:1941}). However, fields with localized structures and Gaussian turbulent fields can exhibit an identical energy spectrum, while particles moving within will have quite different transport properties.

The various efforts to numerically simulate the transport of relativistic particles in turbulent magnetic fields started with the pioneering work by \cite{giacalone-jokipii:1999}. This and subsequent work \cite[e.g.,][and references therein]{casse_lemoine_pelletier:2001,qin-etal-2002, tautz-2010, tautz-dosch:2013, laitinen-etal-2012, reichherzer-etal-2020} were based on a superposition of Fourier modes aimed at testing for which regimes the transport remains diffusive (as opposed to, e.g., subdiffusive perpendicular to the magnetic guide field) and, if so, also aimed at the structure of the diffusion tensor. Most of these simulations made use of specific turbulence spectra
like isotropic, slab, or slab/two-dimensional ones and employed a homogeneous guide field (see e.g. \cite{dundovic-pezzi-blasi-etal:2020} where a detailed comparison of these approaches with respect to parallel and perpendicular diffusion is presented). 
Simulations using turbulent electromagnetic fields obtained from direct numerical simulations of the MHD equations have been conducted  by e.g.\ \cite{cohet-marcowith:2016} and 
\cite{wisniewski-spanier-kissmann:2012}. These direct approaches have the disadvantage that only very limited Reynolds numbers can be achieved.
Studies of particle transport in anisotropic turbulence \citep{pommois-zimbardo-veltri:2007} have already shown that, in addition to diffusion, superdiffusion as well as subdiffusion can occur subject to the strength of the anisotropy. 
Several papers address the effects of intermittency on particle transport.
\cite{alouanibibi-leroux:2014} introduced an 
intermittency model in which the deviation from a Gaussian PDF is generated by modifying the amplitudes of the plane wave modes with a $q$-Gaussian statistic. 
\cite{pucci-malara-etal:2016} applied the so-called $p$-model \citep{meneveau-sreenivasan:1987} to generate intermittent turbulent fields. One conclusion from this study was that intermittency mainly increases parallel transport.
\cite{shukurov-snodin-etal:2017} introduced an intermittent magnetic field by solving the induction equation with a given velocity field. Their conclusion was that intermittency has a profound effect on the diffusion tensor, especially for energies $E \lesssim 10^{10}$\,GeV.
See also \citep{tharakkal-snodin-etal:2023} for the effect of magnetic traps on the spatial distribution of cosmic rays.
The conjecture that intermittency most likely has no impact on the diffusion of ultra high-energy particles is confirmed in a laser-driven plasma experiment
\citep{chen-bott-etal:2020}.
\cite{durrive-lesaffre-ferriere:2020} presented a method for generating magnetic turbulence by generalizing an approach from fluid dynamics \citep{pereira-garban-chevillard:2016}. This method is based on a generalized Biot-Savart kernel that takes into account the stretching of the vorticity encoded in the Cauchy-Green tensor. Fourier methods are used to calculate the integral.

To summarize the situation, extremely sophisticated methods (see also \cite{juneja-lathrop-etal:1994,biferale-bofetta-etal:1998,robitaille_StatisticalModelFilamentary_2020,muzy:2019,luebke-friedrich-grauer:2023,luebke-effenberger-etal:2024,durrive_SwiftGeneratorThreedimensional_2022,maci-keppens-bacchini:2024})
were developed in the last decades to generate synthetic turbulent fields obeying given energy spectra but more importantly a given scaling of higher order structure functions as expected in hydrodynamics \citep{she-leveque:1994,yakhot:2006} or MHD turbulence \citep{grauer-krug-marliani:1994}.
However, these methods do not provide the appropriate coherent structures such as shocks, vortex tubes or current sheets that seem necessary to describe the transport of cosmic particles in turbulent fields.
In somewhat technical terms, even the infinite set of higher order two-point structure function exponents do not describe the necessary correlations to characterize coherent structures: \textit{nearly singular structures introduce non-trivial structure function scaling but the converse does not hold.}

An attempt to introduce coherent structures in the synthetic fields was made based on an idea by \citet{rosales-meneveau:2008} in the work of \citet{subedi-chhiber-etal:2014}. An improved version with scale dependent advection was introduced in \citet{luebke-effenberger-etal:2024}.
Even though the model includes intermittency and successfully generates pronounced coherent structures, it does only slightly improve the transport of charged test particles compared to the results from direct numerical simulations of MHD.
Additionally, to create effective synthetic fields for studying cosmic ray transport and modulation, it is essential to understand the key interactions between cosmic rays and turbulent structures that lead in the desired transport behavior.
Some of these interactions are fast transport in coherent structures (\cite{shukurov-snodin-etal:2017,hu-lazarian-etal:2022}), magnetic moment scattering due to sharp field line bends (\cite{lemoine:2023,kempski-drummond-etal:2023}) and magnetic mirroring (\cite{lazarian-xu:2021,zhang-xu:2023}).

In this paper we take a different approach to attack the problem 
of having a quantitative accurate stochastic model for the complex motions of cosmic rays in a turbulent magnetic field. We apply a state-of-the-art generative Diffusion Model to generate Lagrangian trajectories of cosmic rays propagated by the Lorentz force. 
Diffusion Models have become apparent in the last couple of years, outperforming other generative models like Generative Adversarial Networks \citep{dhariwal-nichol:2021, atmos15010060}. They were originally developed for image synthesis \citep{ho-jain-etal:2020}, but have found great use in various other domains, including medical imaging \citep{oezbey-dalmaz-etal:2023}, audio semantic communication \citep{grassucci-marinoni-etal:2024}, molecule design \citep{bilodeau-jin-etal:2022}, etc. Their main strength lies in the ability to learn any probability distribution for a given training set and generate samples with high diversity and fidelity. 
Recently, \citet{li2024synthetic, li2024generative} successfully employed a diffusion model to generate Lagrangian trajectories in 3D hydrodynamic turbulence, accurately reproducing all pertinent statistical benchmarks, including multi-scaling non-Gaussian fluctuations and cross-correlation between the three velocity components.
This paper expands upon that work from a magnetohydrodynamics perspective, incorporating a highly turbulent magnetic field obtained from direct numerical simulations for the generation of cosmic ray trajectories.
The goal is to provide a first quantitatively correct  {\it black-box} stochastic modeling of this intricate phenomenon for charged particles over a range of different energies.
This is intended as a first step towards more and more expressive generative models, able to extrapolate to energies outside of the provided training data.
This is also a first step towards building of sophisticated non-linear interpolation and imputation techniques, capable to refill gaps in observed temporal series  \citep{cressie1990origins, williams2006gaussian, luebke-friedrich-grauer:2023, li2024stochastic}.
The grand long-term challenge is to learn in a reduced latent variable space an effective interpretable interaction potential able to generate the Lagrangian trajectories in terms of a few physically-motivated  hyper-parameters, as recently proposed for wavelet score-based generative models ~\citep{guth2022wavelet, morel2024scale, lempereur2024hierarchic}.
See also \citep{palade:2024} for predicting cosmic ray diffusion coefficients with the aid of a neural network.

The paper starts by describing the methods in Section~\ref{sec:methods}, consisting of the baseline test particle simulation, training of  and sampling from the generative diffusion model, and the statistics employed for evaluating the synthetic trajectories.
Section~\ref{sec:results} discusses the obtained results, consisting of non-Gaussian anomalous scaling of short-time velocity increments, long-time integrated spatial transport, as well as the curvature and torsion of the gyro-center motion.
Additionally, trajectories obtained from two synthetic turbulent fields are included to the comparison.
Finally, Section~\ref{sec:discussion} concludes the paper by providing a summary and further outlook.

\section{Methods}\label{sec:methods}
\subsection{Charged Particle Motion in MHD Turbulence}\label{sec:data}
In this paper we consider incompressible magnetohydrodynamic flows with constant density  \citep{biskamp:2003, goedbloed-keppens-poedts:2019} in a uniform periodic cube goverened by
\begin{subequations}
\begin{gather}
    \pdv{\vb{u}}{t}+ \vb{u}\cdot\nabla\vb{u} - \vb{b}\cdot\nabla\vb{b}=-\nabla p +\nu\nabla^2\vb{u}+\vb{f},\quad  \nabla\cdot\vb{u}=0 \\
    \pdv{\vb{b}}{t} + \vb{u}\cdot\nabla\vb{b} -\vb{b}\cdot\nabla\vb{u} = \eta\nabla^2\vb{b},\quad \nabla\cdot\vb{b}=0.
\end{gather}
\end{subequations}
Incompressible magnetohydrodynamic turbulence is characterized by the magnetic Prandtl number $\mathcal{P}_m = \frac{\nu}{\eta}$ and the associated kinetic and magnetic Reynolds numbers. We do not discuss features of magnetohydrodynamic turbulence here but refer to the excellent review by \cite{schekochihin:2022}
where the current picture of the important features in the understanding of the MHD turbulence are summarized. We performed simulations of MHD turbulence using the pseudo-spectral code \textit{SpecDyn} which is tailored for modern HPC architectures \citep{wilbert_NumericalInvestigationFlow_2022, wilbert_ImplementationApplicationPseudospectral_2023}. 
Simulations were performed with a resolution of $N^3 = 1024^3$ and a  magnetic Prandtl number $\mathcal{P}_m = 1$ with~$\nu=\eta=1.2 \times 10^{-3}$. The kinetic and magnetic Reynolds based on the Taylor scale are $\mathcal{R}_{\lambda}=439$ and $\mathcal{R}_{\lambda, m}=94$. The velocity field was driven by a random force on low Fourier modes $1\le k \le 3 $ using a forcing scheme proposed in \cite{alvelius_RandomForcingThreedimensional_1999} which is characterized by a low noise level in the time evolution of the turbulence bulk quantities. Forcing is only applied to the velocity field, the magnetic field is generated by the dynamo action in three dimensions.

Test particles are propagated through static snapshots of the MHD simulation by solving the Newton-Lorentz equations
\begin{equation}
    \dv{\vb{X}}{t}=\vb{V},\quad
    \dv{\vb{V}}{t}=\alpha\,\vb{V}\times\vb{b}\big(\vb{X}(t)\big),
    \label{eq:lorentz}
\end{equation}
where the dimensionless parameter $\alpha=t_0\omega_g=l_0/r_g$ is proportional to the particle's gyro-frequency $\omega_g=b_\mathrm{rms}|q|/m$ and the inverse of the particle's maximal gyro-radius $r_g=v_0/\omega_g$, where $l_0$ and $t_0$ denote the typical length and time scale in the system.
An important feature of the particle dynamics is that the particle's kinetic energy $mv_0^2/2$ is conserved.
The parameter $\alpha$ gives the coupling strength between the test particle and the underlying magnetic field, which is related to the particle's energy (low energy particles are coupled strongly to the field, while high energy particles are coupled weakly).

Equation~(\ref{eq:lorentz}) is solved with a Boris scheme \citep{boris:1970}, which conserves the kinetic energy up to machine precision.
A fixed timestep is chosen as $\Delta t=10^{-2}\times2\pi/\alpha$, such that a typical gyration $T_g={2\pi}/{\alpha}$ is resolved with 100 steps.
In total we simulate $96\,000$ particles per $\alpha$ in ten statistically independent snapshots of the MHD simulation with periodic boundary conditions and record their trajectories consisting of spatial and velocity components $\vb{X}(t)$ and $\vb{V}(t)$, as well as the magnetic field $\vb{b}\big(\vb{X}(t)\big)$.
The trajectories have a length of $1\,000$ typical gyrations, resulting in $100\,000$ data points per trajectory and are allowed to repeatedly cross through the domain.
For $\alpha=512$ the typical gyro-radius is comparable to the grid size, particles closely follow magnetic fieldlines and are susceptible to confinement in coherent structures.
For $\alpha=32$ the typical gyro-radius is much larger, so the particles experience a coarse average of the magnetic field, resulting in trajectories akin to a random walk.
Additionally we consider $\alpha=128$ as an intermediate case.
Also note that temporal statistics of charged particles in static snapshots are really biased spatial statistics of the underlying magnetic field.

\subsection{Generative Diffusion Model}
\begin{figure}
    \centering
    \includegraphics[width=0.8\textwidth]{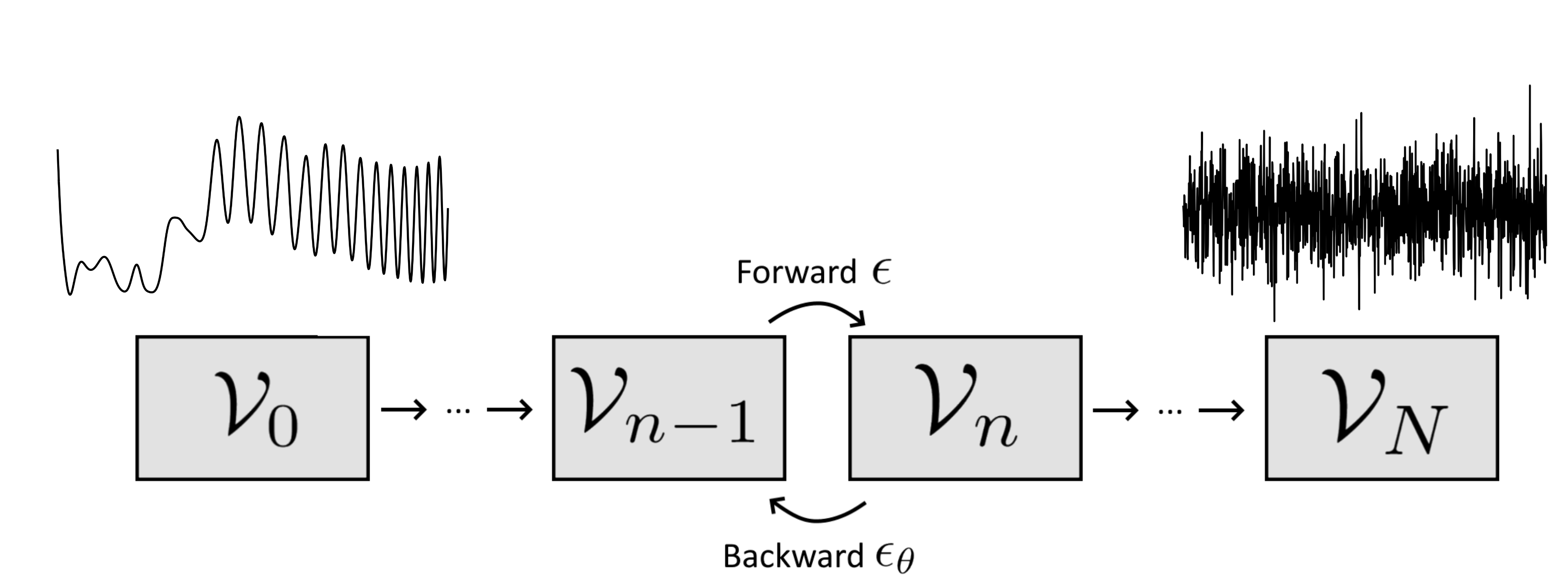}
    \caption{Schematic illustration of the Diffusion Model (DM) process.}
    \label{fig:markov}
\end{figure}

For generating the velocity trajectories $\mathcal{V}=\{\vb{V}(t_k)\mid t_k=k\,\Delta t,\,k=0,\cdots,N-1\}$ at $N$ discrete times, we employ a Generative Diffusion Model \citep{ho-jain-etal:2020, nichol2021improved}. 
Diffusion models (DM) consist of a forward process and a backward process, as schematically depicted in Figure~\ref{fig:markov}.
The forward process is a Markov chain that iteratively adds Gaussian noise $\epsilon\sim\mathcal{N}(\vb{0},\mathrm{Id})$ to training data $\mathcal{V}_0$ via
\begin{equation}
    \mathcal{V}_n=\sqrt{1-\beta_n}\,\mathcal{V}_{n-1}+\sqrt{\beta_n}\,\epsilon,
    \label{eq:fwd}
\end{equation}
such that the distribution of $\mathcal{V}_n$ converges to $\mathcal{N}(\vb{0},\mathrm{Id})$ as $n\to\infty$. The variances $\beta_n$ are predetermined and follow a linear noise schedule (see Table~\ref{tab:hyperparameters}).
Since we are working with Gaussian noise, Equation~(\ref{eq:fwd}) can be written in terms of the initial and final trajectories as
\begin{equation}
    \mathcal{V}_n=\sqrt{\Bar{\alpha}_n}\,\mathcal{V}_0+\sqrt{1-\Bar{\alpha}_n}\,\epsilon,
    \label{eq:fwd0}
\end{equation}
with $\Bar{\alpha}_n=\prod_{i=1}^{n}(1-\beta_i)$,
which removes the need to compute all intermediate steps $\{\mathcal{V}_1,\cdots,\mathcal{V}_{n-1}\}$.
The transition probability distributions for Equations~(\ref{eq:fwd}) and~(\ref{eq:fwd0}) can be found by observing that $\mathcal{V}_n$ is a Gaussian random variable with mean $\sqrt{1-\beta_n}\,\mathcal{V}_{n-1}$ and variance $\beta_n$, or mean $\sqrt{\Bar{\alpha}_n}\,\mathcal{V}_0$ and variance $1-\Bar{\alpha}_n$ as
\begin{align}
    q(\mathcal{V}_n|\mathcal{V}_{n-1})&=\mathcal{N}(\sqrt{1-\beta_n}\,\mathcal{V}_{n-1},\beta_n\,\mathrm{Id}), \\
    q(\mathcal{V}_n|\mathcal{V}_0)&=\mathcal{N}(\sqrt{\Bar{\alpha}_n}\,\mathcal{V}_0,(1-\Bar{\alpha}_n)\,\mathrm{Id}).
\end{align}
The backward process is a reverse Markov chain, modeled by a neural network that learns to undo the forward diffusion steps by sampling from the inverse transition probability
\begin{equation}
    p_\theta(\mathcal{V}_{n-1}|\mathcal{V}_n)=\mathcal{N}\left(\mu_\theta(\mathcal{V}_n,n),\sigma^2_n\,\mathrm{Id}\right),
\end{equation}
where the variance is assumed to be the same as for the forward process, $\sigma_n^2=\beta_n$, and the mean $\mu_\theta(\mathcal{V}_n,n)$ is predicted by the neural network via
\begin{equation}
    \mu_\theta(\mathcal{V}_n,n)=\frac{1}{\sqrt{\Bar{\alpha}_n}}\left(\mathcal{V}_n-\frac{\beta_n}{\sqrt{1-\Bar{\alpha}_n}}\epsilon_\theta(\mathcal{V}_n,n)\right).
\end{equation}
The neural network, parametrized by $\theta$, is trained by minimizing an upper bound of the negative log likelihood, which can be simplified as (see \citet{ho-jain-etal:2020})
\begin{equation}
    \mathcal{L}_\mathrm{simple}=\mathbb{E}_{n,\mathcal{V}_0\sim q(\mathcal{V}_0),\,\epsilon\sim\mathcal{N}(\vb{0},\mathrm{Id})}\left[\|\epsilon-\epsilon_\theta(\mathcal{V}_n,n)\|^2\right],
\end{equation}
where the expected value is taken over $n$ sampled uniformly from 1 to $N$, as well as the training data and Gaussian noise. 
Here $q(\mathcal{V}_0)$ denotes the true data distribution, that we want to approximate with the neural network, which is sampled by choosing a random trajectory from the training data set.
During the training process, the Markov chain is computed directly at step~$n$ by means of Equation~(\ref{eq:fwd0}), which exposes the neural network to a new independent sample from the training data at each training step.
The training algorithm is listed in Algorithm~\ref{Trainingloop}.
Once training is completed, new trajectories can be generated by sampling from the trained model, starting with uncorrelated Gaussian noise and looping through the reverse Markov chain.
The sampling algorithm is listed in Algorithm~\ref{SamplingLoop}.
From a mathematical point of view, the model learns to approximate the probability distribution of all trajectories in the training set. Afterwards, it can sample from that distribution generating new trajectories that obey the same statistics as the training data.
 
\begin{figure}
\begin{algorithm}[H]
  \caption{Training}\label{Trainingloop}
  \begin{algorithmic}[1]
     \State Randomly select a trajectory $\mathcal{V}_0$ from  the training dataset distribution $q(\mathcal{V}_0)$.
     \State  Randomly select an intermediate step $n$ of the Markov chain from a uniform distribution $\{1,..., N\}$.
     \State Create noise according to $\mathbf{\epsilon} \sim \mathcal{N} (\mathbf{0}, \mathrm{Id})$.
     \State Compute the noisy trajectory in the $n$-th step of the Markov chain as $\mathcal{V}_n=\sqrt{\Bar{\alpha}_n}\mathcal{V}_0+\sqrt{1-\Bar{\alpha}_n} \mathbf{\epsilon}$.
     \State Make a gradient descent step with $\nabla_{\theta}||\epsilon-\epsilon_{\theta}(\mathcal{V}_n, n)||^2$ where $\epsilon_{\theta}(\mathcal{V}_n, n)$ is the predicted noise by the network.
     \State Repeat steps 1-5 until converged.
  \end{algorithmic}
\end{algorithm}
\end{figure}

\begin{figure}
\begin{algorithm}[H]
  \caption{Sampling}\label{SamplingLoop}
  \begin{algorithmic}[1]
    \State Create a trajectory that is pure Gaussian noise  $\mathcal{V}_N \sim \mathcal{N} (\mathbf{0}, \mathrm{Id})$.
    \State Loop through the backward Markov chain: 
    \For{$n=N, ..., 1$}
        \If{$n > 1$}
            \State $\epsilon \sim \mathcal{N}(\mathbf{0}, \mathrm{Id})$
        \Else \State $\epsilon = 0$ \Comment{No additional noise in the last step}
        \EndIf
        \State $\mathcal{V}_{n-1}=\mu_\theta(\mathcal{V}_n,n)+\Tilde{\sigma}_n\,\epsilon$
    \EndFor
    \State Return $\mathcal{V}_0$.
  \end{algorithmic}
\end{algorithm}
\end{figure}

The neural network used is based on the work of \citet{NN} and utilizes a U-Net architecture \citep{ronneberger2015u}. This architecture is a fully convolutional neural network consisting of two parts: an encoder and a decoder. Firstly, the training data are progressively down-sampled in the encoder, reducing the dimension of the training data and extracting conceptual features. In the decoder, the network upsamples the data again and learns these features, while also recovering the original dimension. The encoder and decoder are connected by a bottleneck, creating a ``U'' shape, which gives the architecture its name. Each encoder layer is linked to its corresponding layer in the decoder through skip connections, which concatenate encoded features with the decoded data. This design enables the model to integrate broad contextual information from the encoder with precise spatial details from earlier layers.
Details on network architecture and hyperparameters used are provided in the Appendix~\ref{app:hyperparameters}.

As described in Section~\ref{sec:data}, our training dataset consists of $96\,000$ trajectories, each sampled at $100\,000$ grid points, for three normalized gyro-frequencies $\alpha=32$, 128 and 512.
Since our primary interest lies in small-scale Lagrangian properties, we train the model on the velocity $\vb{V}(t)$, which is normalized such that $\norm{\vb{V}(t)}=1$.
To train the model more effectively, the velocity data is transformed from Cartesian to spherical coordinates:
\begin{align*}
  r &=\norm{\vb{V}(t)}=1,\\
  \theta &=\arccos{V_z(t)},\\ 
  \phi &= \operatorname{sgn}\left(V_y(t)\right)\arccos\left(\frac{V_x(t)}{\sqrt{V_x(t)^2+V_y(t)^2}}\right),
\end{align*}
which replaces the three velocity components $(V_x, V_y, V_z)$ with two angular quantities ($\theta$ and $\phi$), thereby reducing the dimensionality of the training data and ensuring the conservation of energy in the generated trajectories.
Furthermore, each trajectory is divided into multiple segments with a length that captures all relevant physics, namely $1\,024$, $2\,048$ and $8\,192$ grid points for $\alpha=32$, $128$, and $512$ and corresponding to $(10.24, 20.48, 81.92)\,T_g$ units, respectively. Here we have used $T_g = 100\ \text{grid points}$ as given by the timestep $\Delta t=0.01\,T_g$ (see Section~\ref{sec:data}).
It is important to emphasize that this does not compromise the statistical information, as the generalized flatness remains constant at scales greater than these shortened lengths. As a result, this approach increases the amount of training data and reduces the training time significantly.

\subsection{Statistics}
Since cosmic ray trajectories are Lagrangian, they can be quantitatively described by several statistical quantities~\citep{frisch1995turbulence, PRL2003, PRL2004, PRL2008, ARFM2009}. The most notable of these is the Lagrangian structure function,
\begin{equation}
    S_{\tau}^{(p)}=\langle (V_i(t+\tau)-V_i(t))^p \rangle=\langle(\delta_\tau V_i)^p\rangle,
    \label{eq:sf}
\end{equation}
where $i\in\{x, y, z\}$ indicates the three velocity components and $\tau$ is the time increment.
Due to isotropy, the dependence on the direction $i$ is left out.  In this equation, the average is taken over both time and over all trajectories. In regular hydrodynamical turbulence, global scaling power laws can be observed by measuring $S_{\tau}^{(p)}$ in the inertial range. A closely related quantity is the generalized flatness, defined as
\begin{equation}
    F^{(p)}_{\tau}=\frac{S_{\tau}^{(p)}}{\Big(S_{\tau}^{(2)}\Big)^{p/2}},
    \label{eq:flat}
\end{equation}
which describes the behavior of the tails of the probability distribution of the velocity increments $\delta_\tau V_i$.
A Gaussian distribution has  $F^{(4)}_{\tau} = 3$, whereas smaller and larger values indicate respectively sub-Gaussian and super-Gaussian tails.
The intermittency of the flow can be investigated by observing changes of the flatness as a function of $\tau$ \citep{structurefunction}.
This indicates that the tails of the increment distribution vary across scales, i.e.~that they exhibit anomalous scaling.

In order to assess the transport properties of the cosmic ray trajectories, which are most interesting for space and astrophysical applications, we investigate the mean squared displacement (MSD)
\begin{equation}
    \langle\Delta X^2_\tau\rangle=\langle\|\vb{X}(t+\tau)-\vb{X}(t)\|^2\rangle,
    \label{eq:msd}
\end{equation}
where the spatial trajectory is recovered by integrating the velocity signal
\begin{equation}
    \vb{X}(t)=\int_0^t\vb{V}(t)\dd{t}.
    \label{eq:spatial-trajectory}
\end{equation}
The process at hand can be characterized by how the MSD scales with time, i.e.~$\langle\Delta X^2_\tau\rangle\sim\tau^a$, with $a=1$ indicates usual Brownian diffusion, $a>1$ indicates superdiffusion and $a<1$ indicates subdiffusion \citep{METZLER2000,aerdker-merten-etal:2025}.
Cosmic ray transport is often taken as diffusive on sufficiently long time scales, so a Fokker-Planck equation with diffusion coefficient $D=\lim_{\tau\to\infty}\langle\Delta X^2_\tau\rangle/2\tau$ can be employed \citep{casse_lemoine_pelletier:2001,qin-etal-2002}.

An alternative approach of describing Lagrangian trajectories is by the use of parametric equations to describe three-dimensional spatial curves, which are fully determined by two geometric parameters;
namely the curvature $\kappa$ and the torsion $\theta$ \citep{curvature}. The curvature determines how much the trajectory curves within the plane defined by the tangent and normal vector of the particle, while the torsion determines how much the plane itself changes. They are defined as
\begin{equation}
    \kappa (t) = \frac{\norm{\vb{V}(t)\cross\vb{A}(t)}}{\norm{\vb{V}(t)}^3},
    \label{eq:curv}
\end{equation}
and
\begin{equation}
    \mathcal{\theta}(t) = \frac{\vb{V}(t)\cdot (\vb{A}(t) \cross \dot{\vb{A}}(t))}{\kappa^2 \norm{\vb{V}(t)}^6},
    \label{eq:torsion}
\end{equation}
with $\vb{A}(t)=\frac{\dd\vb{V}(t)}{\dd t}$ and $\dot{\vb{A}}(t)=\frac{\dd\vb{A}(t)}{\dd t}$. 
We focus our treatment primarily on the curvature, since it offers more useful insight than the torsion due to its intuitive interpretation.
Further, since a charged particle in a magnetic field undergoes helical motion, its curvature is given by its instantaneous gyro-radius $r_g(t)$ and pitch angle $\vartheta(t)$.
As shown in Appendix~\ref{app:curvature}, we have $\kappa(t)
=\alpha\sin\vartheta(t)$
where the instantaneous quantities are related to the particle parameter $\alpha$ via $\alpha=\sin\vartheta(t)/r_g(t)$,
and thus obtain the mean curvature by integrating over the distribution of all pitch angles
\begin{equation}
    \langle\kappa\rangle=\alpha\int_0^\pi\sin^2\vartheta\,\dd\vartheta=\frac{\pi}{2}\alpha.
\end{equation}
When estimating the distribution of $\kappa(t)$ over the raw trajectories, we obtain
a pronounced peak at $\langle\kappa\rangle$ followed by a steep decay, irregardless of the model which generated the particular trajectories.
Instead, it is more interesting to look at the curvature distribution of the gyro-center motion.
We find that low-pass filtering the velocity signal with a filter size of $1.5\,T_g$ (which corresponds to 150 grid points) gives a good approximation
\begin{equation}
    \Bar{\vb{V}}(t)=\int_{t-1.5T_g}^t\vb{V}(t')\dd{t'}.
\end{equation}
Thus, $\kappa(t)$ computed on $\Bar{\vb{V}}(t)$ supplements the previously discussed Lagrangian statistics, which is most interesting for sub-gyro-period scales, by geometric insight on gyro-period scales.


\section{Results}\label{sec:results}
\begin{figure}
    \centering
    \includegraphics[width=\linewidth]{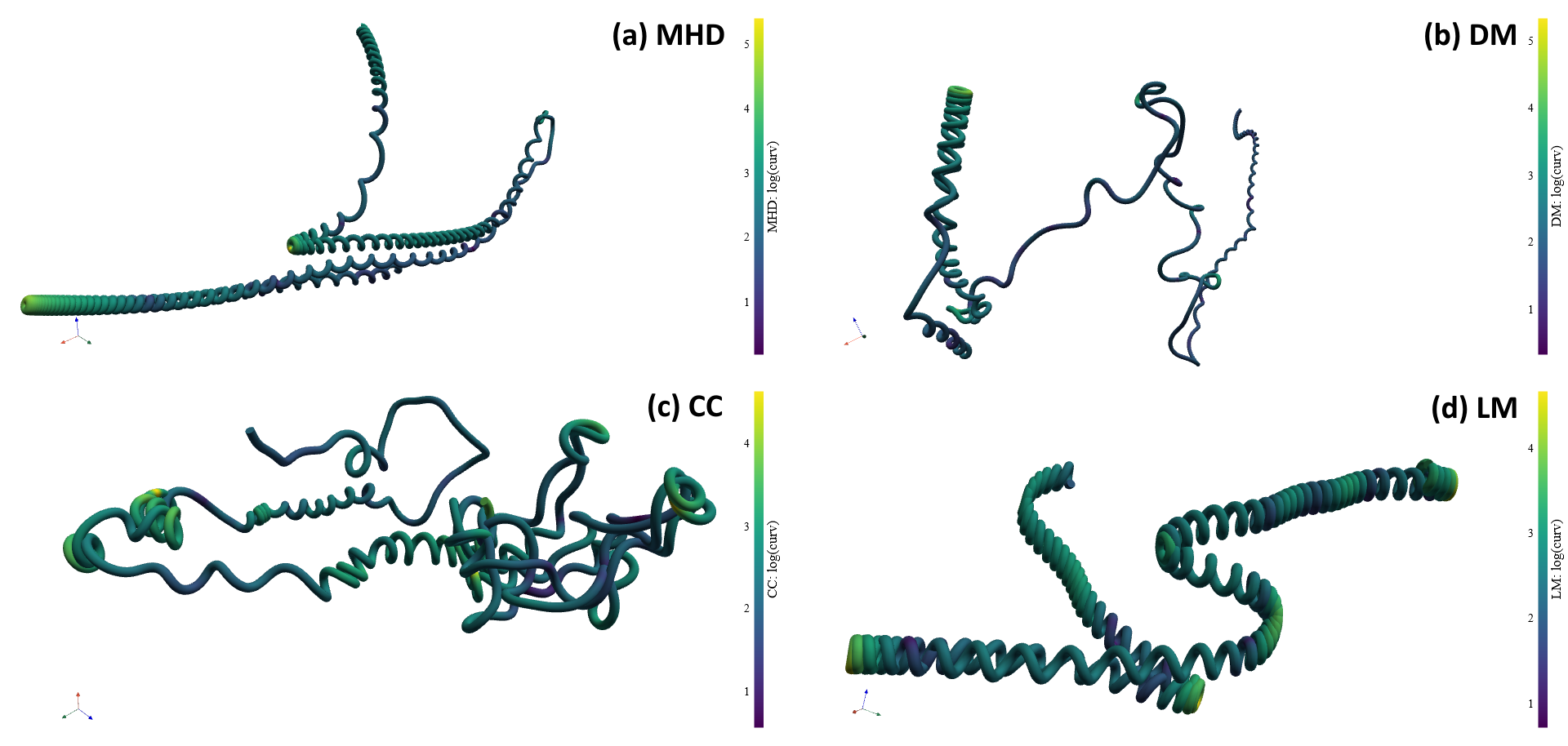}
    \caption{Example trajectories for different models of charged particles in magnetic turbulence, colored by the curvature of the gyro-center motion.
    The cases (a), (c) and (d) are obtained by solving the Newton-Lorentz equations in a turbulent magnetohydrodynamics (MHD) snapshot, a multifractal continuous cascade (CC) field and a structured Lagrangian mapping (LM) field. The case (b) is generated by our generative diffusion model (DM).}
    \label{fig:example-trajectories}
\end{figure}

We now discuss the statistics, which are presented in the previous section, for four models of charged particle trajectories in magnetic turbulence.
Firstly, the trajectories obtained by numerically integrating Equation~(\ref{eq:lorentz}) in static snapshots from our magnetohydrodynamics (MHD) simulation serves as the baseline.
Secondly, the trajectories generated by our generative diffusion model (DM) are of our primary interest, because we can assess the performance of the trained neural network by their statistics.
We apply an additional low-pass filtering with a window of 3 grid points (i.e.~$\Delta\tau=0.03\,T_g$) to the DM trajectories, to account for the model struggling with small-scale smoothness.
Additionally, we add trajectories from two models for synthetic turbulence \citep{luebke-effenberger-etal:2024} to the discussion:
the continuous cascade (CC) model provides a random multi-fractal unstructured vector field, and the Lagrangian mapping (LM) model incorporates simple advective shock-like coherent structures.
In both cases the trajectories are also obtained by numerically integrating Equation~(\ref{eq:lorentz}), where the respective random vector field is taken as the turbulent magnetic field $\vb{b}(\vb{x})$.
For all four cases we consider $10\,000$ independent vector-valued velocity signals with lengths $8192$, $2048$ and $1024$ for~$\alpha=512$, 128 and 32, respectively.
An example trajectory with $\alpha=512$ for each model is shown in Figure~\ref{fig:example-trajectories}.

\subsection{Anomalous Scaling}
\begin{figure}
    \centering
    \includegraphics[width=0.75\linewidth]{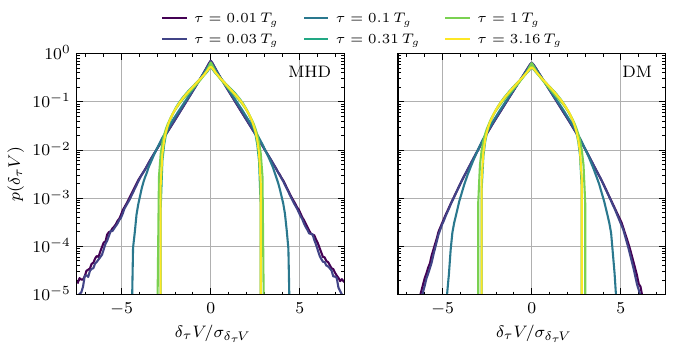}
    \caption{Distributions of normalized increments $\delta_\tau V/\sigma_{\delta_\tau V}$ for different time lags $\tau$. Shown are the MHD and DM cases with $\alpha=512$.
    Power-law tails for small $\tau$ with strong non-Gaussianity, and confined sub-Gaussian behavior for large $\tau$ due to~$V_i\in[-1,1]$.}
    \label{fig:incr-distr}
\end{figure}

We approach the analysis of intermittency in the velocity signals by looking at the distribution of velocity increments~$\delta_\tau V_i=V_i(t+\tau)-V_i(t)$ normalized by their standard deviation $\sigma_{\delta_\tau V_i}$, and comparing increasing lags $\tau$.
From the estimated distributions plotted in Figure~\ref{fig:incr-distr}, we observe strong exponential tails for small $\tau$, and a quick convergence to the confined sub-Gaussian behavior for $\tau\gtrsim T_g$.
This sub-Gaussian behavior is due to the energy conservation of Equation~(\ref{eq:lorentz}), such that $V_i\in [-1, 1]$ and consequently $\delta_\tau V_i\in[-2, 2]$.
We also note, that the DM struggles to match the exponential tails for small $\tau$ and large $\delta_\tau V$.
We investigate this anomalous scaling behavior more quantitatively by looking at scale-dependent statistics, specifically the second-order structure function $S_\tau^{(2)}$ given in Equation~(\ref{eq:sf}) and the fourth-order flatness $F_\tau^{(4)}=S_\tau^{(4)}/\big(S_\tau^{(2)}\big)^2$ given in Equation~(\ref{eq:flat}).

\begin{figure}
    \centering
    \includegraphics[width=\linewidth]{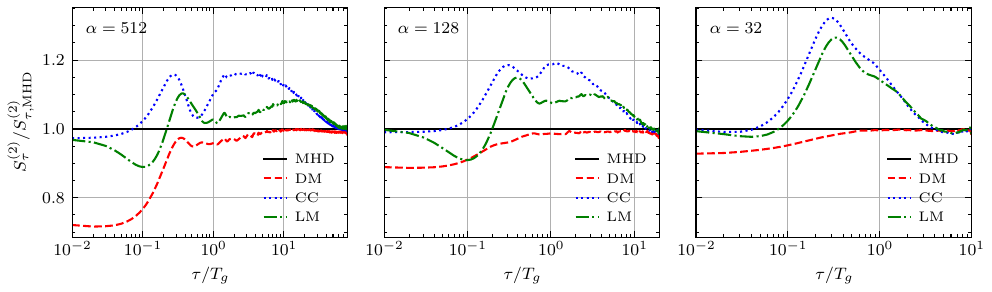}
    \caption{Second-order velocity structure functions $S^{(2)}_\tau$ of particle trajectories, compensated by the MHD case to highlight differences.}
    \label{fig:s2}
\end{figure}

The second-order structure functions, compensated by the MHD baseline case to highlight the differences, are plotted in Figure~\ref{fig:s2}.
The previously observed struggle of the DM to match the exponential tails for small lags $\tau$ appears again as a relatively large deviation of $S^{(2)}_{\tau,\mathrm{DM}}$ from $S^{(2)}_{\tau,\mathrm{MHD}}$ for $\alpha=512$; for smaller $\alpha$ values these deviations become less pronounced, and for lags $\tau\gtrsim T_g$ the DM agrees very well with the MHD case.
While the other two models CC and LM  show good agreement for small $\tau$, pronounced deviations arise for $\tau\sim 0.1\,T_g,\cdots,10\,T_g$, which are present for all considered $\alpha$ values.
This is likely due to the MHD, CC and LM trajectories being generated by solving Equation~(\ref{eq:lorentz}), so in the small $\tau$ regime the $S^{(2)}_\tau$-behavior is governed by basic physical dynamics of charged particle, while for larger $\tau$ differences in the magnetic field models take effect.

The scale-dependent flatness is plotted in Figure~\ref{eq:flat}.
The MHD case exhibits an increased flatness $F^{(4)}_\tau\approx 5.6$ for small lags $\tau$ corresponding to the exponential tails, which quickly goes to $F^{(4)}_\tau\approx 3$ for $\tau\sim 0.5\,T_g$, where it exhibits an oscillation caused by the regular gyro-motion of the particles, before converging to its asymptotic value $F^{(4)}_\tau\approx 2.4$ corresponding to the confined sub-Gaussian tails.
This behavior is consistent for all considered $\alpha$ values, except for the oscillation at $\tau\sim 0.5\,T_g$, which becomes more washed-out as $\alpha$ decreases.
This is due to larger gyro-radii of low-$\alpha$ particles, so fine magnetic structures cannot be followed any longer with a regular gyro-motion, and the trajectories resemble more and more unstructured random walks.
The DM struggles again somewhat at small $\tau$, but exhibits for $\tau\gtrsim0.1\,T_g$ very well with the MHD case, which indicates that the neural network robustly learned the gyro-center motion guided by the MHD structure.
Meanwhile, the CC and LM models perform rather bad under this metric.
The CC case appears attenuated compared to the MHD case with smaller values of $F^{(4)}_\tau$ throughout small and intermediate lags $\tau$, as well as exhibiting a shorter oscillation at $\tau\sim 0.5\,T_g$.
On the other hand, the LM case exhibits extreme flatness values of $F^{(4)}_\tau\approx 12$ for small lags $\tau$, which indicates extreme velocity fluctuations over short time scales.

\begin{figure}
    \centering
    \includegraphics[width=\linewidth]{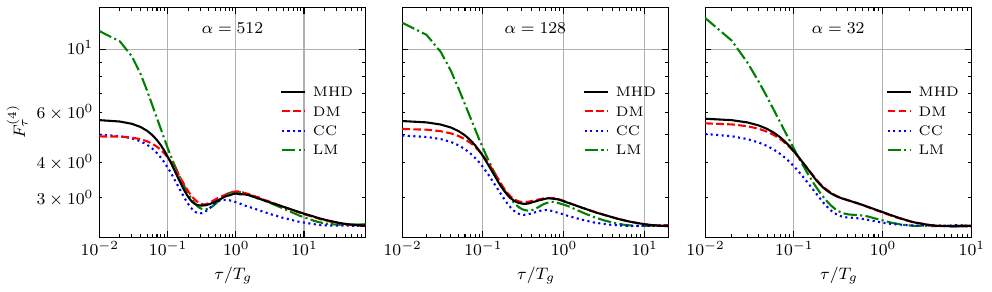}
    \caption{Fourth-order velocity flatness $F^{(4)}_\tau$ of particle trajectories, indicating the deviation from the Gaussian case $F^{(4)}_\tau=3$.}
    \label{fig:f4}
\end{figure}

\subsection{Transport Behavior}
\begin{figure}
    \centering
    \includegraphics[width=\linewidth]{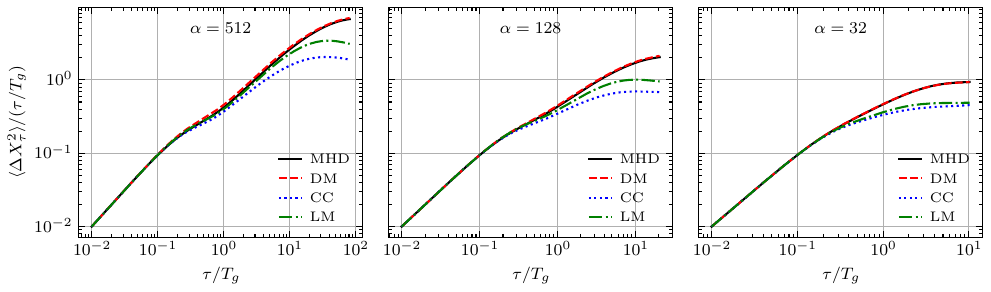}
    \caption{Mean squared displacement as an indicator for the transport properties of charged particles. Compensated by the time $\tau$ to distinguish between ballistic and diffusive behavior.}
    \label{fig:msd}
\end{figure}

The mean squared displacement of spatial trajectories $\vb{X}(t)$ as given by Equation~(\ref{eq:msd}), which indicates the spatial spread of a population of particles after a given time, is plotted in Figure~\ref{fig:msd}.
The plots are compensated by the time $\tau$ to easily identify the respective transport regimes.
For short times $\tau<0.1\,T_g$ all models exhibit identical unperturbed ballistic transport $\langle\Delta X^2_\tau\rangle\sim\tau$.
Previously observed differences of the velocity statistics do not come into effect at this level of consideration, because by recovering the spatial trajectory $\vb{X}(t)$ from the velocity signal $\vb{V}(t)$ by Equation~(\ref{eq:spatial-trajectory}) and averaging according to Equation~(\ref{eq:msd}), the effects of the intermittent short-time velocity fluctuations are washed out.
However, at times $\tau\gtrsim T_g$ the effect of the gyro-center motion being influenced by the underlying magnetic field structures becomes apparent again, as observed for the velocity statistics.
Observing the behavior of $\langle\Delta X^2_\tau\rangle$ for DM trajectories substantiates the previous assertion that the DM neural network robustly learns the gyro-center motion as governed by coherent structures in MHD turbulence.
In comparison, for an entirely unstructured field (CC) or for a field with weak coherent structures (LM), charged particles do not achieve the same mean squared displacements, and thus they underestimate the diffusion coefficients observed in MHD turbulence.

\subsection{Curvature of the Gyro-center Motion}
After discussing short-time velocity statistics and long-time transport behavior, we now take a closer look the geometry of the trajectories.
The motion of charged particles in turbulent magnetic fields can be viewed as a superposition of a small-scale ($\tau\lesssim T_g$) gyro-motion and a large-scale ($\tau\gtrsim T_g$) gyro-center motion guided by and scattered off structures in the turbulent field.
As discussed in section~\ref{sec:methods}, the small-scale gyro-motion dominates the curvature distribution of the unfiltered trajectories, which peaks at $\langle\kappa\rangle$ followed by a steep decay.
On the other hand, the curvature distribution of the gyro-center motion, which can be approximated by low-pass filtering the velocity signal $\vb{V}(t)$ with a window size $\Delta\tau\sim1.5\,T_g$, can give some insight into how the particle trajectories are shaped by the underlying turbulent field.

We note that several previous works discussed curvature of Lagrangian trajectories in Navier-Stokes turbulence~\citep{braun_delillo_eckhardt:2006,xu_CurvatureLagrangianTrajectories_2007,curvature,bentkamp_StatisticalGeometryMaterial_2022,qi_FoldingDynamicsIts_2023} and fieldlines in magnetic turbulence~\citep{yang_RoleMagneticField_2019,yuen_CurvatureMagneticField_2020,bandyopadhyay_SituMeasurementCurvature_2020}.
Additionally, \citet{lemoine:2023} and \citet{kempski-drummond-etal:2023} recently pointed out how cosmic ray transport is affected by the magnetic fieldline curvature.
In contrast, we consider here the curvature of charged particle trajectories in turbulent fields, which --- to the best of our knowledge --- has not been investigated before.

\begin{figure}
    \centering
    \includegraphics[width=\linewidth]{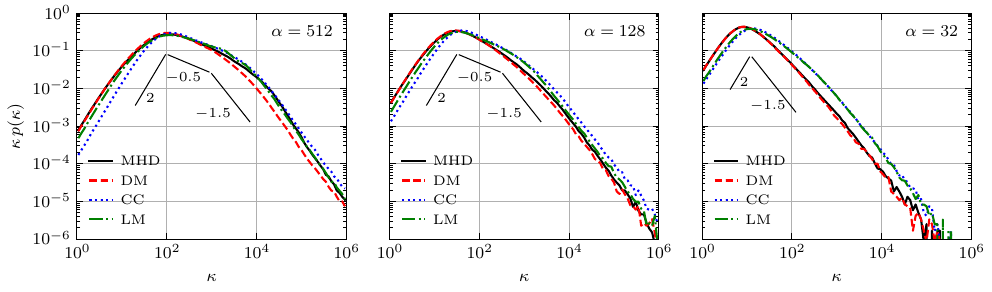}
    \caption{Distributions of curvature of gyro-center trajectory, by averaging the velocity signals $\vb{V}(t)$ with a window of $1.5\,T_g$.
    The gyro-center motion exhibits an asymptotic tail with slope $-2.5$, and, for large $\alpha$, an intermediate regime with slope $-1.5$.
    The low-curvature tails scale with slope $1$.
    The distributions are compensated with $\kappa$ for visualization purposes, the indicated slopes take this compensation into account.}
    \label{fig:curv-filtered}
\end{figure}

The curvature distributions of the gyro-averaged particle trajectories are plotted in Figure~\ref{fig:curv-filtered},
where we observe the asymptotic scalings $p(\kappa)\underset{\kappa\to0}{\sim}\kappa$ and $p(\kappa)\underset{\kappa\to\infty}{\sim}\kappa^{-2.5}$ for all turbulence models and all values of $\alpha$.
Additionally, we observe the intermediate scaling $p(\kappa)\sim\kappa^{-1.5}$ for $\alpha=512$ and 128.
The asymptotic scalings $\kappa$ and $\kappa^{-2.5}$ have been consistently observed in fluid and magnetic turbulence as reported in the aforementioned references, and can be explained by writing the curvature as $\kappa=a_\perp/V$ with the perpendicular acceleration~$a_\perp=\big\|\frac{\Bar{\vb{V}}}{V}\times\dot{\Bar{\vb{V}}}\big\|$ and the velocity magnitude~$V=\|\Bar{\vb{V}}\|$ (cf.~Equation~(\ref{eq:curv})).
By taking either $a_\perp$ or $V$ as a random variable with a given distribution and the other variable as a function of the first one, we derive the observed scaling behaviors following \citet{curvature} as shown in Appendix~\ref{app:curvature}.
The results are summarized in Table~\ref{tab:kappa-scalings}.
\citet{xu_CurvatureLagrangianTrajectories_2007} and \citet{bandyopadhyay_SituMeasurementCurvature_2020}
present an alternative approach based on Taylor-expansions of $\chi_\mathrm{dof}^2$-distributions for $a_\perp$ (with $\mathrm{dof}=2$) and $V$ (with $\mathrm{dof}=3$), while keeping the respective other variable constant.

Applying these results to, e.g., the case $\alpha=512$, we observe a low-curvature tail $\sim\kappa$ due to on average constant~${V\sim\order{1}}$ and small Gaussian~$a_\perp$, which corresponds to regular motion along coherent fieldlines.
Further, at intermediate curvatures the scaling $\sim\kappa^{-1.5}$ implies on average constant $a_\perp$ and intermediate $V$ roughly distributed according to a uniform distribution.
The values of $V$ are bounded, since the raw components are bounded as $V_i\in[-1,1]$ and the low-pass filtering is done with a finite window size.
Finally, the high curvatures, which scale as $\sim\kappa^{-2.5}$, are caused by small, roughly Gaussian~$V$ and finite~$a_\perp$.
This regime entails sharp, sudden turns of the trajectories, and, specifically for large $\alpha$, direction-reversals due to magnetic mirror events.

\begin{figure}
    \centering
    \includegraphics[width=0.8\linewidth]{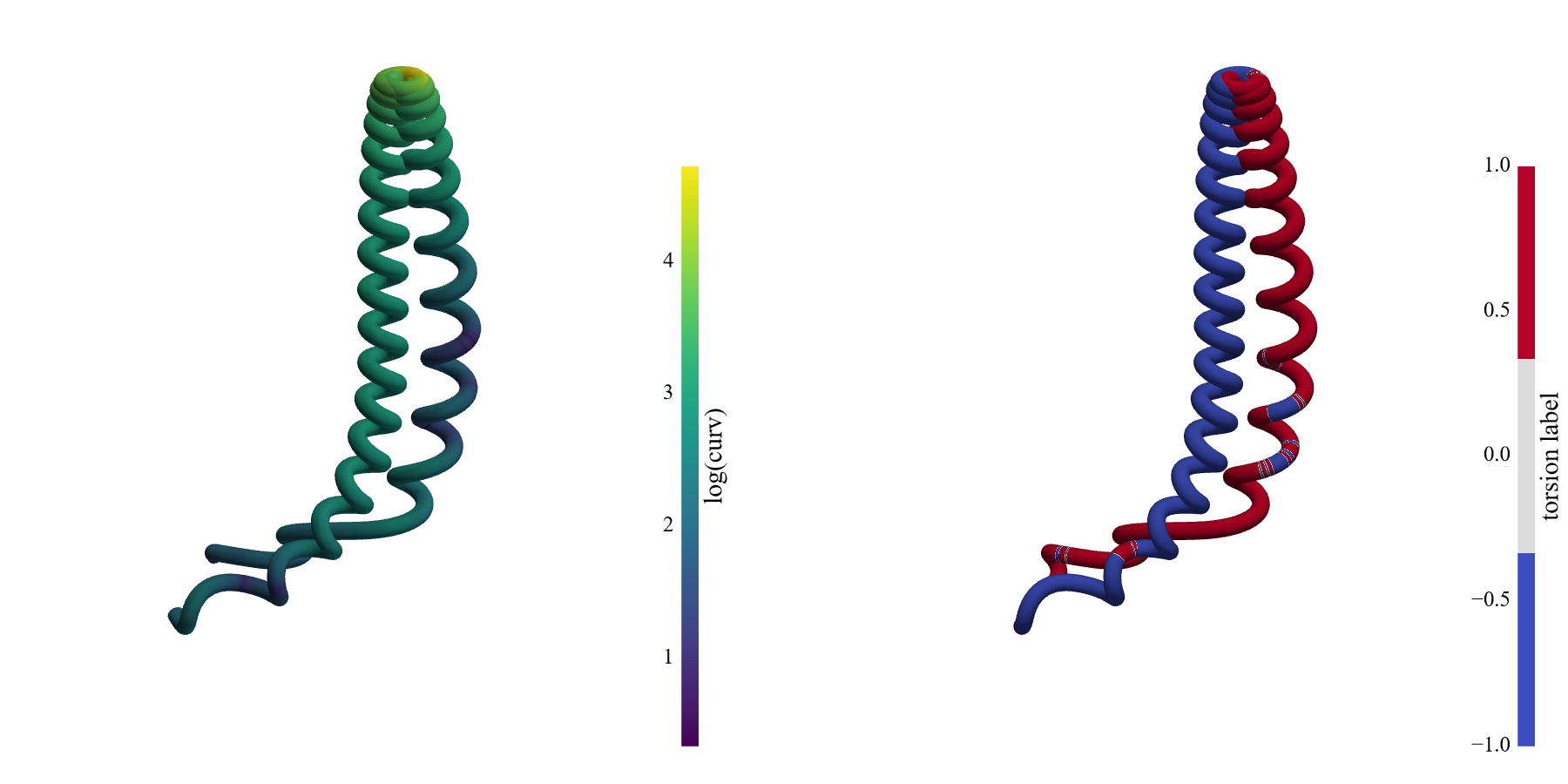}
    \caption{Example of a mirror event in a DM-generated trajectory.
    The event is characterized by high curvature and a change of the handedness of the helix, as indicated by the sign of the torsion.
    The curvature and torsion are computed for the gyro-center motion.}
    \label{fig:mirror-example}
\end{figure}

Magnetic mirrors are characterized by slowly converging fieldlines, where the adiabatic conservation of the magnetic moment of trapped particles causes a steady reversal $V_\parallel$.
These events also change the handedness of the helical particle motion, as indicated by the sign of the torsion, as depicted in Figure~\ref{fig:mirror-example}.

\begin{figure}
    \centering
    \includegraphics[width=.45\linewidth]{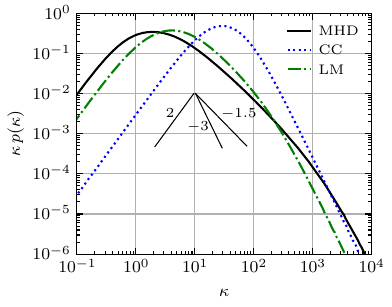}
    \caption{Distribution of fieldline curvature in turbulent vector fields. The $-2.5$ asymptotic scaling is only present in the MHD case, while the CC case is characterized by a $-4$ slope.
    All low-curvature tails scale also with slope $1$.
    The distributions are again compensated with $\kappa$, which is taken into account by the indicated slopes.}
    \label{fig:curv-fields}
\end{figure}

Lastly, we take a look at the fieldline curvature distributions of the underlying turbulent fields for the MHD, CC and LM models, which are shown in Figure~\ref{fig:curv-fields}.
The above explanations of the asymptotic scaling behavior apply in the Eulerian case as well, where the curvature is given by~$\kappa=a_\perp/b^2$ with the magnetic field strength~$b=\|\vb{b}\|$ and the perpendicular acceleration~${a_\perp=\big\|\frac{\vb{b}}{b}\times\big(\vb{b}\cdot\nabla\vb{b}\big)\big\|}$.
While the MHD field exhibits the expected $\kappa^{-2.5}$ scaling, the CC field instead scales as $\kappa^{-4}$, which is also observed for Gaussian random fields. This implies that $a_\perp$ is not constant on average, but rather behaves as $a_\perp\sim b$ according to Table~\ref{tab:kappa-scalings}.
The asymptotic particle curvature scaling due to on average constant perpendicular acceleration $a_{\perp,\mathrm{gyro-center}}$ is likely a physical property of the gyro-center motion and independent from the asymptotic fieldline curvature scaling.
Firstly, particles tend to scatter away from their current fieldline upon encountering sharp fieldline reversals as discussed by \citet{lemoine:2023} and \citet{kempski-drummond-etal:2023},
secondly, sharp particle reversals due to magnetic mirroring tend to occur in low-curvature areas of the field,
and thirdly, the $\kappa_\mathrm{gyro-center}^{-5/2}$ scaling persists even in the CC and LM fields with distinct high $\kappa_\mathrm{fieldline}$ behavior.

\begin{table}
    \centering
    \begin{tabular}{cccc}
        \toprule
         $V$ & $a_\perp$ & Limit & Scaling  \\
         \midrule
         const. & $\chi_2(a_\perp)$ & $\kappa\to0$ & $\kappa^1$ \\
         $\chi_3(V)$ & const. & $\kappa\to\infty$ & $\kappa^{-2.5}$ \\
         $\chi_3(V)$ & $V$ & $\kappa\to\infty$ & $\kappa^{-4}$ \\
         $\mathcal{U}(V)$ & const. & $\kappa\to\infty$ & $\kappa^{-1.5}$ \\
         \bottomrule
    \end{tabular}
    \caption{Scaling regimes of curvature distributions for $\kappa=a_\perp/V^2$. The variables $a_\perp$ and $V$ are either given as a random variable with a prescribed distribution or as a function of the other variable.
    A $\chi_k$-distribution assumes Gaussian one-point statistics with $k$ degrees of freedom, while a uniform distribution assumes boundedness of the variable.
    See Appendix~\ref{app:curvature} for the derivations.}
    \label{tab:kappa-scalings}
\end{table}

\section{Conclusions and Outlook}\label{sec:discussion}
\textit{Conclusion:}
We presented a new approach to model trajectories of energetic charged particles in turbulent magnetic fields, which utilizes a generative diffusion model.
To this end we trained neural networks on velocity signals obtained from integrating the Newton-Lorentz equation in static snapshots of an MHD turbulence simulation, and investigated the generated trajectories with respect to short-time velocity statistics, long-time transport behavior and geometric characteristics by means of the curvature of the gyro-center motion.
Additionally, we compared the results with trajectories obtained from two state-of-the-art models for random synthetic turbulence, which include multifractal statistics and simple coherent structures.

In summary, the DM excels at accurately reproducing the physical properties of the baseline MHD trajectories such as velocity statistics at intermediate and long times, the long-time transport behavior which is of special interest to space and astrophysical applications, and the geometry of the gyro-center motion, which reflects, e.g., magnetic mirror events.
However, we were so far unable to generalize the model to unseen particle energies (interpolation and extrapolation of $\alpha$), which requires separately trained networks for each case.
On the other hand, while synthetic turbulence models clearly struggle to achieve high accuracy of physical properties (more work is needed, but potential is clearly seen), they present an easier method to estimate cosmic ray behavior for parameters outside of the DNS-accessible domain.

\textit{Outlook:}
A separate model was trained for each value of $\alpha$ encoding the energy of cosmic ray particles. The task now is to train a model that makes different values of $\alpha$ directly accessible. A similar situation exists with respect to the background magnetic field and the associated anisotropic turbulence. In this paper, only homogeneous isotropic turbulence was analyzed. A future model must make different background fields directly accessible for the synthesis of cosmic ray trajectories.
In the long term, it would be desirable to be able to generate longer trajectories from the existing data. This could be achieved through a combination of stable diffusion and established methods for generating synthetic fields on large scales with approximate Gaussian statistics.
An extension to the case of the acceleration of cosmic particles in the presence of electric fields, both through shocks, reconnection and turbulence, would be especially useful.
From this description it becomes clear that the contribution in this paper represents only a first step in connection with stable diffusion in the space and astrophysical context.
All of this is necessary to address the grand long-term challenge of learning an effective, interpretable interaction potential in a reduced latent variable space. This potential should be able to generate Lagrangian trajectories with a few physically motivated hyperparameters, as recently proposed for wavelet score based generative models. \\
\\
J.L. and R.G. like to thank M. Wilbert for sharing data from his MHD simulations. We also like to thank H. Fichtner and F. Effenberger for useful discussions.
J.L and R.G. are supported by the Deutsche Forschungsgemeinschaft (DFG, German Research Foundation) within the Collaborative Research Center SFB1491.
J.L. and R.G. gratefully acknowledge the Gauss Centre for Supercomputing e.V. (\url{www.gauss-centre.eu}) for funding this project by providing computing time on the SuperMUC-NG at Leibniz Supercomputing Centre (\url{www.lrz.de}) and through the John von Neumann Institute for Computing (NIC) on the GCS Supercomputers JUWELS at Jülich Supercomputing Centre (JSC).
This work was supported by the European Research Council (ERC) under the European Union’s Horizon 2020 research and innovation programme Smart-TURB (Grant Agreement No. 882340).

\appendix
\section{Network architecture and hyperparameters}
\label{app:hyperparameters}
\begin{figure}
    \centering
    \includegraphics[width=\textwidth]{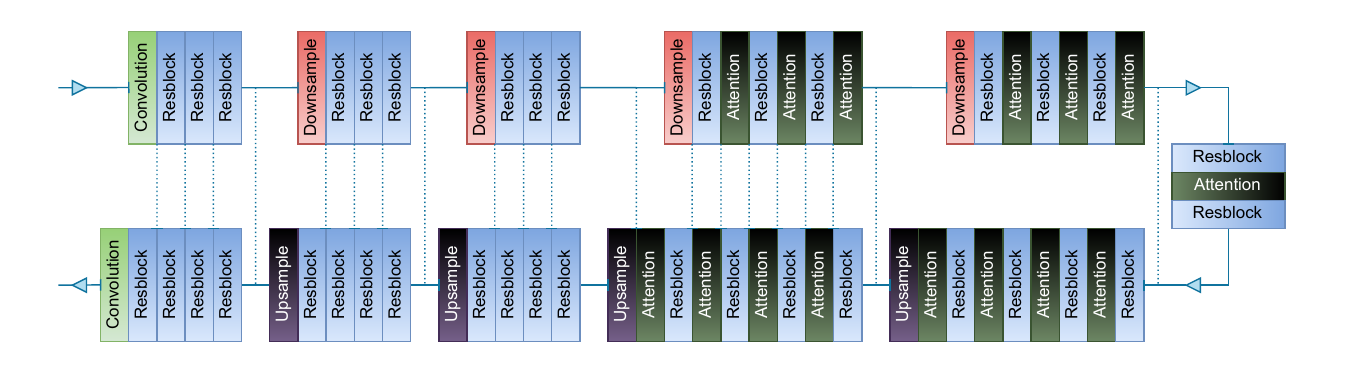}
    \caption{Example diagram of the employed UNet architecture with 5 layers. The upper row corresponds to the encoder, the lower row to the decoder, and the vertical part to the bottleneck. The dashed lines represent the connecting paths where the data is concatenated.}
    \label{fig:Unet}
\end{figure}

A sketch of the employed architecture is given in Figure~\ref{fig:Unet}, here consisting of five layers. 
The architecture consists of several blocks, namely downsample, upsample, residual- and attention blocks. The downsample and upsample blocks are responsible for the respective consecutive doubling and halving of the resolution through each layer. The residual blocks enhance information flow across the neural network and effectively reduce the vanishing gradient problem that often arises in deep neural networks. Finally, the attention block, as the name suggests, allows the network to focus on the most important features of the data while ignoring less important ones.
During the training process of the neural network, we studied the influence several hyperparameters, namely  the number of diffusion steps, layers and initial channels (i.e.~the number of feature maps produced by convolutional filters), and subsequently fine-tuned their values for each particle energy.
Their final values are given in Table~\ref{tab:tuned}.
Additional fixed hyperparameters are given in Table~\ref{tab:hyperparameters}.

\begin{table}
\centering
\begin{tabular}{lccc}
     \toprule
     Hyperparameter & $\alpha=512$ & $\alpha=128$ & $\alpha=32$ \\
     \midrule
     Number of diffusion steps & {1000} & {1000} & {1000} \\
     Number of UNet layers & {6} & {10} & {9} \\
     Number of initial channels & {64} & {64} & {64} \\
     \bottomrule
\end{tabular}
\caption{Fine-tuned hyperparameters for each particle $\alpha$.}
\label{tab:tuned}
\end{table}

\begin{table}
\centering
\begin{tabular}{lcl}
\toprule
Hyperparameter & Value & Explanation\\
\midrule
Depth                & 3            & number of residual blocks per layer \\
Heads                & 1            & number of attention heads per attention block \\
Attention Resolution & 250, 125     & length of the downsampled data vector, for which attention blocks are active \\
Batch size           & 64           & number of trajectories simultaneously fed to the network during each training step \\
Learning rate        & $10^{-4}$    & step size of the gradient descent algorithm \\
Noise schedule       & Linear       & noise variance $\beta_n$ for each diffusion step, linearly increasing from $\beta_1=10^{-4}$ to $\beta_N=0.02$ \\
\bottomrule
\end{tabular}
\caption{Fixed hyperparameters.}
\label{tab:hyperparameters}
\end{table}

\section{Network performance}
\label{app:performance}
Investigating the performance of the neural network when certain hyperparameters are varied is often a main interest
in machine learning studies. Typically, it is true that increasing the model size leads to better results, but this is
accompanied by an increased training time. It is therefore important to optimize this trade-off. This work investigated the
number of channels, diffusion steps, and layers of the network. It was found that an increase in the number of channels and 
diffusion steps did not yield a significant improvement, while greatly affecting the training time. Additionally, multiple
models were trained where the provided training data consisted not only of the two angular quantities $\theta$ and $\phi$,
but also of the pitch angle, thus containing information on the underlying magnetic field. Once again, no significant 
improvement was found by including this additional quantity. However, the number of employed layers in the network did 
affect the accuracy of the produced trajectories. Generally, a minimum of three layers were required to produce adequate
results. By increasing the number of layers further, no significant advancements were made with regard to the global error,
although often there existed a trade-off of the accuracy on different time scales. 

\begin{figure}
    \centering
    \includegraphics[width=\textwidth]{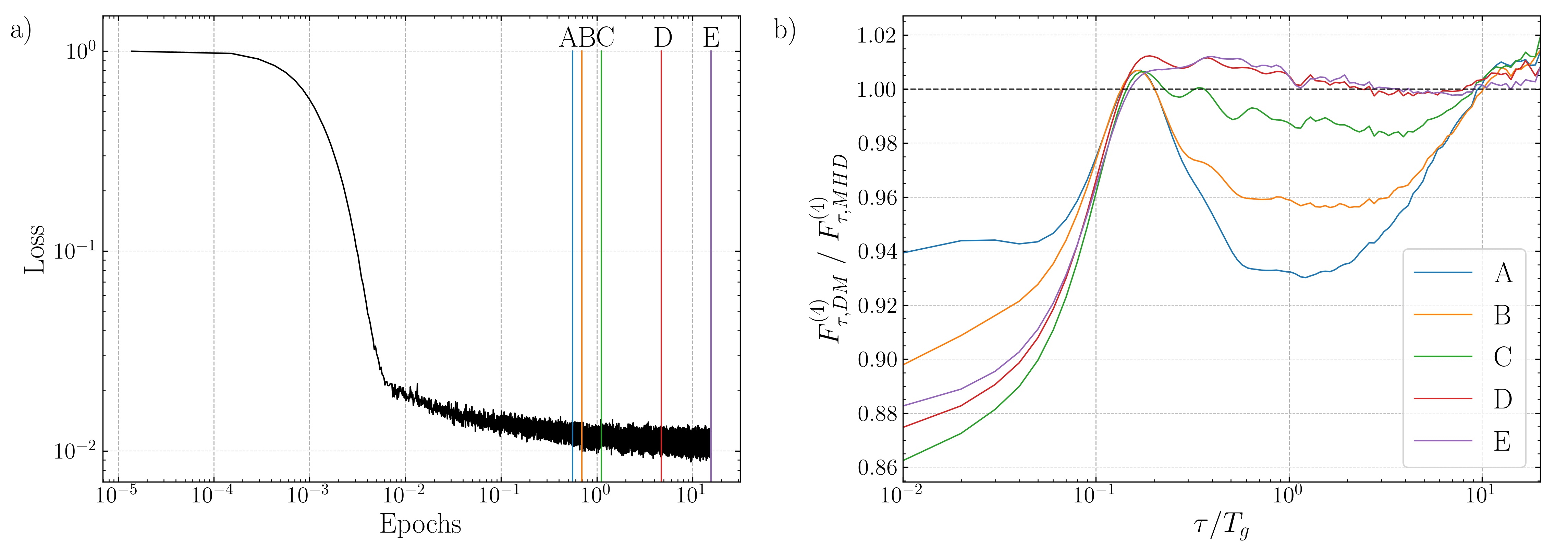}
    \caption{a) Training loss function against epochs. b) Fourth-order flatness $F^{(4)}_\tau$ of the particle trajectories generated by the diffusion model at 5 different training times (A-E, highlighted in the training loss function), compensated by the MHD case to highlight
differences. }
    \label{fig:LossConvergence}
\end{figure}

\begin{figure}
    \centering
    \includegraphics[width=0.7\textwidth]{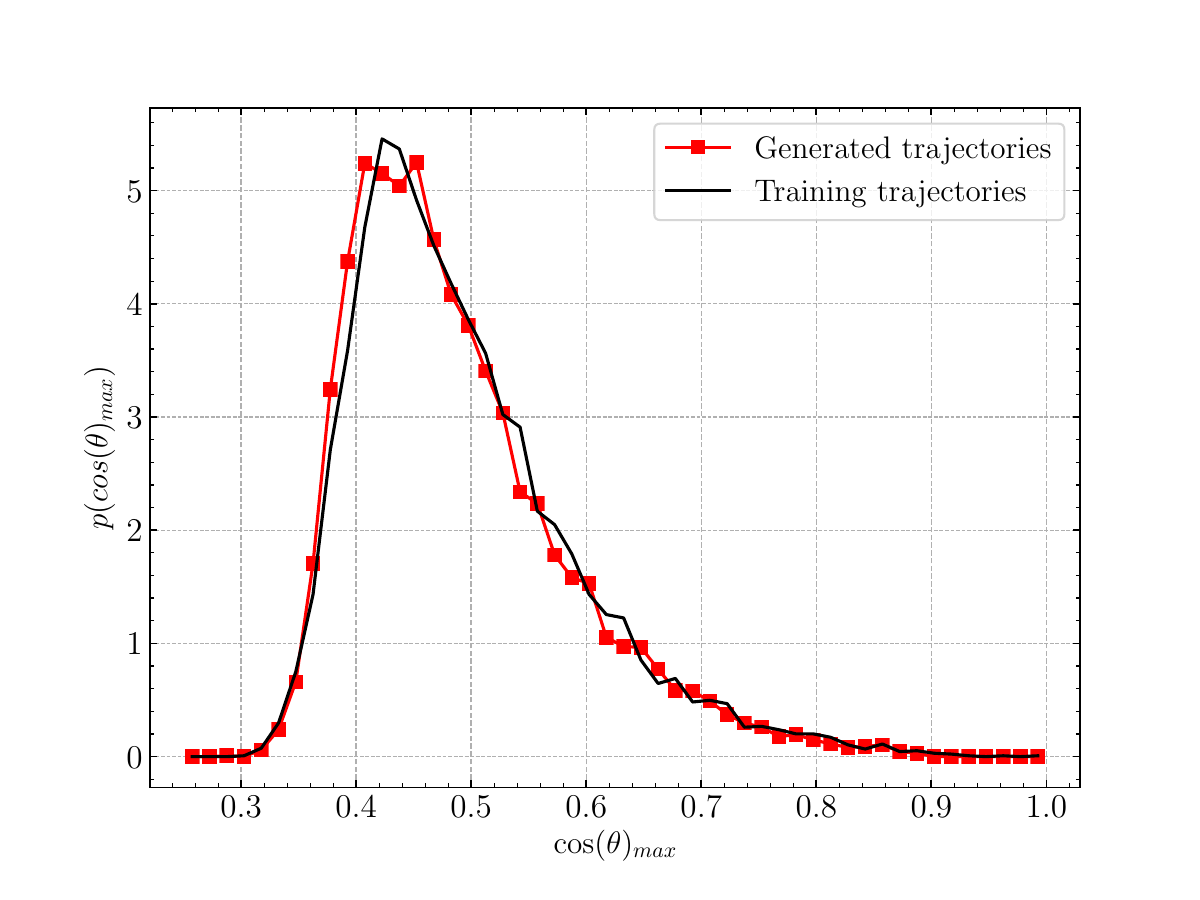}
    \caption{Distribution of the cosine similarity of the closest training trajectory for a generated sample (red) and training trajectory (black) for $\alpha=128$.}
    \label{fig:cosinesim}
\end{figure}

Figure \ref{fig:LossConvergence}a) shows the training loss function plotted against the number of epochs for $\alpha=128$. Although the loss function quickly convergences, it is important that training is continued to ensure the accuracy of the results. This is illustrated in Figure \ref{fig:LossConvergence}b) where the fourth-order flatness of the trajectories generated by the diffusion model, normalized by the MHD training data, is plotted at five different training times (denoted by A-E in the loss function). Initially, the accuracy of the results varies greatly (A-C), but requires increasingly longer training for notable effects (D-E). Additionally, it can be seen that the model prioritizes the accuracy of the intermediate and long scales, where a clear converging trend towards $F^{(4)}_{\tau, DM}/F^{(4)}_{\tau, MHD}=1$ is present. In contrast, no such behavior is seen for the small time scales.
Finally, it is important to verify the generalization of the neural network by checking whether the model generates new trajectories without memorization. For each sample, the closest trajectory in the training data set is found by computing the cosine similarity, defined as \begin{equation}
    \cos{\theta} = \frac{ \vb{V_i} \cdot \vb{V_j}}{\|\vb{V_i}\| \|\vb{V_j}\|} = \frac{\sum_{t=0}^{t_{f}} \vb{V_{i}}(t) \vb{V_{j}}(t)}{\sqrt{\sum_{t=0}^{t_{f}} \vb{V_i}(t)^2}\sqrt{\sum_{t=0}^{t_{f}} \vb{V_j}(t)^2}}.
\end{equation}
Figure \ref{fig:cosinesim} 
shows the probability distribution of the cosine similarity between a subset of generated samples and their closest training trajectories, as well as between a subset of training samples and their closest other training trajectories (excluding themselves). If the network had memorized trajectories, a peak at high values would be expected in the distribution of the generated samples. Instead, it can be seen that the distributions agree well, thus showing no evidence of memorization. 

\section{Curvature calculations}
\label{app:curvature}
\subsection{Curvature of helical trajectories}
For a simple helix with radius $a$ and slope $a/b$
\begin{equation}
    \vb{X}(t)=(a\cos t,a\sin t, bt)
\end{equation}
the curvature is given by
\begin{equation}
    \kappa=\frac{a}{a^2+b^2}.
\end{equation}
For the trajectory of a charged particle with parameter $\alpha$ and pitch angle $\vartheta$, we have $a=\frac{\sin\vartheta}{\alpha}$ and $b=\frac{\cos\vartheta}{\alpha}$, and thus find
\begin{equation}
    \kappa=\alpha\sin\vartheta.
\end{equation}
The average curvature is then found by assuming a uniform distribution of the pitch angle cosine and computing the integral
\begin{equation}
    \langle\kappa\rangle=\alpha\int_{-1}^{+1}\sin\vartheta\,\dd{\cos\vartheta}=\alpha\int_0^{2\pi}\sin^2\vartheta=\frac{\pi}{2}\alpha.
\end{equation}

\subsection{Asymptotic scaling of curvature distributions}
In order to derive the asymptotic scaling of the curvature distribution $p(\kappa)$, we write the curvature as
\begin{equation}
    \kappa=\frac{a_\perp}{w^2},
\end{equation}
with $\vb{w}=\vb{V}(t)$ and $a_\perp=\left\|\frac{\vb{V}(t)}{V(t)}\times\dot{\vb{V}}(t)\right\|$ for Lagrangian particle trajectories and $\vb{w}=\vb{b}(\vb{x})$ and $a_\perp=\left\|\frac{\vb{b}(\vb{x})}{b(\vb{x})}\times \big(\vb{b}(\vb{x})\cdot\nabla\vb{b}(\vb{x})\big)\right\|$ for Eulerian vector fields.
The high-curvature tail is given by low $w$ values.
We observe $a_\perp\sim\mathrm{const.}$ for low and intermediate values of $V(t)$ of the particle gyro-center trajectories (lowpass filtered) and for intermediate values of $b(\vb{x})$ in the MHD case.
Further we observe $a_\perp\sim w$ for the lowest values of $b(\vb{x})$ in the MHD case and for low and intermediate values of $b(\vb{x})$ in the CC case, as well as for Gaussian random fields.
Taking the one-point distribution of $w_i$ as Gaussian, $w=\|\vb{w}\|$ is distributed according to a $\chi$-distribution with $k=3$ degrees of freedom
\begin{equation}
    p(w)\sim w^2e^{-\frac{w^2}{2}}.
\end{equation}

Then, by making use of properties of the $\delta$-distribution and dropping any constants, we derive the scaling $p(\kappa)\sim\kappa^{-5/2}$ for $a_\perp\sim\mathrm{const.}$
\begin{align*}
    p(\kappa)&=\int p(w)\,\delta\!\left(\kappa-\frac{a_\perp}{w^2}\right)\dd{w} \\
    &\sim\int p(w)\,\delta\!\left(\kappa-w^{-2}\right)\dd{w} \\
    &\sim\int p(w)\,\frac{\kappa^{-3/2}}{2}\,\delta\!\left(w-\kappa^{-1/2}\right)\dd{w} \\
    &=\frac{1}{2}\,p(w=\kappa^{-1/2})\,\kappa^{-3/2} \\
    &=\frac{1}{2}\kappa^{-5/2}e^{-\frac{1}{2\kappa}}
    \underset{\kappa\to\infty}{\sim}\kappa^{-5/2}.
\end{align*}
Similarly, we derive the scaling $p(\kappa)\sim\kappa^{-4}$ for $a_\perp\sim w$
\begin{align*}
    p(\kappa)&=\int p(w)\,\delta\!\left(\kappa-\frac{a_\perp}{w^2}\right)\dd{w} \\
    &\sim\int p(w)\,\delta\!\left(\kappa-w^{-1}\right)\dd{w} \\
    &\sim\int p(w)\,\kappa^{-2}\,\delta\!\left(w-\kappa^{-1}\right)\dd{w} \\
    &=p(w=\kappa^{-1})\,\kappa^{-2} \\
    &=\kappa^{-4}e^{-\frac{1}{2\kappa}}
    \underset{\kappa\to\infty}{\sim}\kappa^{-4}.
\end{align*}
Finally, if $w$ is a bounded variable with a distribution $p(w)\sim\mathrm{const.}$ and we take $a_\perp\sim\mathrm{const.}$, repeating the above calculation yields
\begin{equation*}
    p(\kappa)\sim\kappa^{-3/2}.
\end{equation*}

On the other hand, in the low $\kappa$ limit we can take $w\sim\mathrm{const.}$ and take Gaussian distribution for the acceleration, which implies a $\chi$-distribution with $k=2$ degrees of freedom for $a_\perp$, i.e,
\begin{equation}
    p(a_\perp)\sim a_\perp e^{-\frac{a_\perp^2}{2}}.
\end{equation}
Then we can derive the scaling $p(\kappa)\sim\kappa^1$ via
\begin{align*}
    p(\kappa)&=\int p(a_\perp)\,\delta\!\left(\kappa-\frac{a_\perp}{w^2}\right)\dd{a_\perp} \\
    &\sim\int p(a_\perp)\,\delta\!\left(\kappa-a_\perp\right)\dd{a_\perp} \\
    &=p(a_\perp=\kappa) \\
    &=\kappa e^{-\frac{\kappa^2}{2}} \underset{\kappa\to0}{\sim}\kappa.
\end{align*}

\bibliography{references}
\bibliographystyle{aasjournal.bst}
\end{document}